\documentstyle[prl,aps,twocolumn]{revtex}
\begin{document}
\draft

\twocolumn[\hsize\textwidth\columnwidth\hsize\csname@twocolumnfalse\endcsname
\title{Fundamental Forces as Gauge Theories}
\author{Heui-Seol Roh\thanks{e-mail: hroh@nature.skku.ac.kr}}
\address{BK21 Physics Research Division, Department of Physics, Sung Kyun Kwan University, Suwon 440-746, Republic of Korea}
\date{\today}
\maketitle

\begin{abstract}
This study proposes that all the known fundamental forces including gravity may be
described by local gauge theories. Gravitational, electroweak, and strong interactions
on length scales from $10^{-33}$ cm to $10^{28}$ cm are systematically discussed from
the unified gauge theory point of view toward a ultimate theory for fundamental
forces. New concepts such as dynamical spontaneous symmetry breaking, gauge group
hierarchy, coupling constant hierarchy, effective coupling constant hierarchy,
cosmological constant, massive gauge bosons, massless gauge bosons, quantum
weakdynamics, analogy between quantum weakdynamics and quantum chromodynamics, a
possible gauge theory for the universe expansion, the relation between time and gauge
boson mass, other quantum tests, etc. are briefly reported based on experiments.
\end{abstract}

]
\narrowtext

The longstanding problem in physics is the unification of gravitation theory and the
other gauge theories for electroweak and strong interactions, which has attracted a
great deal of work. Two fundamental theories, general relativity and quantum gauge
theory, can together cover the energy regime from the elementary particle to the
universe in describing the behavior of particles. However, the problem of unifying
gravity and the other fundamental forces in nature is that general relativity theory
is difficult to formulate as a renormalizable gauge theory. All the known covariant or
canonical quantization methods are not yet successful in quantizing the gravitational
field, even though they work well for the other physical fields.
General relativity or standard big bang theory based on general relativity has outstanding
problems: the singularity, cosmological constant or vacuum energy,
flat universe, baryon asymmetry, horizon problem, the large scale
homogeneity and isotropy of the universe, dark matter, galaxy
formation, discrepancy between astrophysical age and Hubble age, etc.
On the other hand, according to the recent experiments, BUMERANG-98 and MAXIMA-1 \cite{Jaff},
the universe is flat and there exists repulsive force represented by non-zero
vacuum energy, which plays dominant role in the universe expansion.
The outstanding problem as well as the other challenging problems may be resolved systematically in
this paper, which uses local gauge theories to describe all the known fundamental
forces on length scales from the Planck scale $10^{-33}$ cm to the universe scale
$10^{28}$ cm. The present work is mainly restricted to the low, real dimensions of
spacetime without considering supersymmetry and is based on phenomenology below the
Planck scale: Newtonian mechanics, Einstein's general relativity and cosmological
constant \cite{Eins}, quantum gauge theories \cite{Glas,Frit,Geor}, and $\Theta$
vacuum \cite{Hoof2} are taken into account. This paper only briefly reports common
features of gauge theories for fundamental forces, whose details are addressed by
following papers, quantum gravity (QG) as an $SU(N)$ gauge theory for gravitational
interactions \cite{Roh1,Roh11}, quantum weakdynamics (QWD) as an $SU(3)_I$ gauge theory
and Glasshow-Weinberg-Salam (GWS) model for
weak interactions \cite{Roh2,Glas}, and quantum chromodynamics (QCD) and quantum
nucleardynamics (QND) for strong interactions \cite{Frit,Roh3,Roh31}.

Toward quantum gravity as a gauge theory, its features are investigated in quantum
gravity (QG) \cite{Roh1,Roh11}; Newton gravitation constant $G_N$ and cosmological constant
$\Lambda$ are discussed from the viewpoint of quantum gauge theory and quantum tests
supporting this scheme are suggested in addition to its relation to Einstein's general
relativity. QG holds underlying principles such as special relativity, gauge
invariance, and quantum mechanics under the constraint of the flat universe
\begin{equation}
\Omega - 1 = - 10^{-61} .
\end{equation}
The flat universe condition is required in quantum gauge theories and inflation
scenario \cite{Guth} and is verified by the experiments, BUMERANG-98 and MAXIMA-1
\cite{Jaff}. $- 10^{-61}$ represents the $10^{30}$ expansion in one dimension: $N_R =
i/(\Omega -1)^{1/2} \approx 10^{30}$. An $SU(N)$ gauge theory with the $\Theta$ vacuum
term as a trial theory is introduced and dynamical spontaneous symmetry breaking
(DSSB) is adopted even though the exact group for gravity is not known at present. The
Lagrangian density for QG \cite{Roh1,Roh11} is, in four vector notation, given by
\begin{equation}
\label{qchr}
{\cal L}_{QG} = - \frac{1}{2} Tr  G_{\mu \nu} G^{\mu \nu}
+ \sum_{i=1}  \bar \psi_i i \gamma^\mu D_\mu \psi_i
+ \Theta \frac{g_g^2}{16 \pi^2} Tr G^{\mu \nu} \tilde G_{\mu \nu}
\end{equation}
where the bare $\Theta$ term \cite{Hoof2} is added as a single, additional
nonperturbative term to the perturbative Lagrangian density with an $SU(N)$ gauge
invariance. The subscript $i$ stands for the classes of pointlike spinor $\psi$ and
$D_\mu = \partial_\mu - i g_g A_\mu$ for the covariant derivative with the
gravitational coupling constant $g_g$. Particles carry local charges and gauge fields
are denoted by $A_{\mu} = \sum_{a=0} A^a_{\mu} \lambda^a /2$ with matrices
$\lambda^a$, $a = 0,.., (N^2-1)$. The field strength tensor is given by $G_{\mu \nu} =
\partial_\mu A_\nu - \partial_\nu A_\mu - i g_g [A_\mu, A_\nu]$ and $\tilde G_{\mu
\nu}$ is the dual of the field strength tensor. A current anomaly is taken into
account to show DSSB in analogy with the axial current anomaly, which is linked to the
$\Theta$ vacuum in QCD as a gauge theory \cite{Roh3}. Since the $G \tilde G$ term is a
total derivative, it does not affect the perturbative aspects of the theory. The
$\Theta$ term plays important roles on the DSSB of the gauge group and on the
quantization of the matter space and vacuum space. The gauge boson identified as the graviton
is proposed as massive vector boson. Newton gravitation constant is defined as
the effective coupling constant
\begin{equation}
G_N/\sqrt{2}= g_f g_g^2/8 M_G^2 \approx 10^{-38} \ \textup{GeV}^{-2}
\end{equation}
with the gravitational coupling constant $g_g$,
and the gauge boson mass $M_G \simeq M_{Pl} \approx 10^{19}$ GeV in QG \cite{Roh1,Roh11}
just as Fermi weak constant is defined as
$G_F/\sqrt{2}= g_w^2/8 M_W^2 \approx 10^{-5} \ \textup{GeV}^{-2}$ with the weak coupling constant $g_w$
and the intermediate vector boson mass $M_W$ in weak interactions \cite{Glas,Roh2}.
The gravitational factor $g_f$ in gravitational interactions is a factor depending on a gauge group
just as the color factor $c_f$ in strong interactions is a factor depending on color dynamics:
these factors are closely related to charge mixing angles \cite{Roh2,Roh3}.
The effective cosmological constant $\Lambda_e$ representing the vacuum energy density is
related to the gauge boson mass $M_G$ by
\begin{equation}
\Lambda_e = 8 \pi G_N M_G^4
\end{equation}
with $\Lambda_e \sim M_{Pl}^2 \approx 10^{38} \ \textup{GeV}^2$
with $M_G \approx 10^{19}$ GeV at the Planck epoch and $\Lambda_e
\approx 10^{-84} \ \textup{GeV}^2$ with $M_G \approx 10^{-12}$ GeV
at the present epoch. The gauge boson mass $M_G$ decreases through
the condensation of the singlet graviton:
\begin{equation}
M_G^2 = M_{Pl}^2 - g_f g_g^2 \langle \phi \rangle^2 = g_f g_g^2 [A_{0}^2 -
\langle  \phi \rangle^2]
\end{equation}
where $A_{0}$ is the singlet gauge boson with even parity and $\langle  \phi \rangle$ is the
condensation of singlet gauge boson with odd parity. DSSB consists of two simultaneous
mechanisms; the first mechanism is the explicit symmetry breaking
of gauge fields, which is represented by the gravitational coupling constant $g_f
\alpha_g$ and the gravitational fine structure
constant $\alpha_g$, and the second mechanism is the spontaneous
symmetry breaking of gauge fields, which is represented by the condensation of singlet
gauge fields. The effective potential is related to the gauge boson mass by $V_e =
M_G^4 = M_{Pl}^4 - 2 g_f g_g^2 M_{Pl}^2 \langle \phi \rangle^2 + g_f^2 g_g^4 \langle \phi \rangle^4$. It reduces to nearly
zero, $M_G \approx 10^{-12}$ GeV, at the present universe; the present universe
expansion \cite{Hubb} and cosmic microwave background radiation (CMBR) \cite{Penz} are
suggested as conclusive evidences for nearly massless gauge bosons
with the mass $M_G \approx 10^{-12}$ GeV and the number $N_G \approx 10^{91}$ and
for massless gauge bosons (photons) as Nambu-Goldstone bosons \cite{Namb} with the
energy $E_\gamma \approx 10^{-13}$ GeV and the number $N_{t \gamma} \approx 10^{88}$.
For the gauge boson mass $M_G \approx 10^{-12}$ GeV at the present epoch, the
corresponding cosmological constant $\Lambda_0 \simeq 10^{-84} \ \textup{GeV}^2$ or
Hubble constant $H_0 \simeq 10^{-42}$ GeV suggests new types of fundamental forces
responsible for the universe expansion and CMBR: the possible gauge groups for this
process is $SU(3)_R \rightarrow SU(2)_B \times U(1)_A \rightarrow U(1)_g$ with the
extremely strong effective coupling constant $G_S \approx 10^{24} \ \textup{GeV}^{-2}
\approx 10^{61} G_N$.
This is supported by the recent experiments BUMERANG-98 and MAXIMA-1 \cite{Jaff}.
This is consistent with the theoretical, large cosmological
constant at the Planck epoch and the empirical, small cosmological constant at
present. This is also compatible with no detection of gravitons at present since they
are extremely heavy particles with the Planck mass $M_{Pl}$ or extremely high energy
particles with $E \approx M_{Pl}$.

The extremely flat universe $\Omega -1 = - 10^{-61}$ is
defined by the relation $\Omega = (\langle \rho_m \rangle - \Theta \rho_m)/\rho_G$ where $\langle \rho_m \rangle$ is the zero point energy density,
$\rho_m$ is the matter energy density, $\rho_G = M_G^4$ is the vacuum energy density,
and $\Theta$ is the parameter of the $\Theta$ vacuum term representing the surface
term \cite{Hoof2}.
This scheme suggests quantum tests toward resolutions to the inflation \cite{Guth} with the universe size
$R = N_R/M_G$ and the maximum wavevector mode of the vacuum
$N_R = i/(\Omega -1)^{1/2} = (\rho_G/ \Theta \rho_m)^{1/2} \approx 10^{30}$,
dark matter as strongly interacting massive particles (SIMPs) with the mass around $10^{-12}$ GeV
in addition to weakly interacting massive particles (WIMPs) \cite{Schr},
structure formation due to massive gauge bosons and dark matter,
primordial nucleosynthesis, etc. of the universe:
they constitute conclusive evidences of massive gauge bosons at different energy scales.
The condition $\Omega - 1 = - 10^{-61}$ of the flat universe leads to
the $\Theta$ value $\Theta = 10^{-61} \ \rho_G/\rho_m$.
If the conserved matter energy density in the universe is close to the critical density $\rho_m \simeq \rho_c \simeq 10^{-47} \ \textup{GeV}^4$,
the $\Theta$ constant depends on the gauge boson mass $M_G$:
\begin{equation}
\Theta = 10^{-61} \ M_G^4/\rho_c  .
\end{equation}
$\Theta$ values become $\Theta_{Pl} \approx 10^{61}$ at the Planck energy,
$\Theta_{EW} \approx 10^{-4}$ at the weak energy, $\Theta_{QCD} \approx 10^{-12}$ at
the strong energy, and $\Theta_{0} \approx 10^{-61}$ at the present energy of the
universe. This is consistent with the observed results, $\Theta < 10^{-9}$ in the
electric dipole moment of the neutron in strong interactions \cite{Alta} and $\Theta
\simeq 10^{-3}$ in the neutral kaon decay in weak interactions \cite{Chri}. If the
baryon mass density is $\rho_B \equiv \Omega_B \rho_c$, the baryon asymmetry
\begin{math}
\delta_B = N_B/N_{t \gamma} \approx 10^{78}/10^{88} = 10^{-10}
\end{math}
with the baryon number $N_B = B$ at present \cite{Stei0} is consistent with Avogadro's
number $N_A = 6.02 \times 10^{23} \ \textup{mol}^{-1}$ and nuclear matter density $n_n
\approx 1.95 \times 10^{38} \ \textup{cm}^{-3}$. Moreover, lepton asymmetries
$\delta_e = N_e/N_{t \gamma} \approx 10^{81}/10^{88} = 10^{-7}$ for electrons,
$\delta_\mu = N_\mu/N_{t \gamma} \approx 10^{79}/10^{88} = 10^{-9}$ for muons, and
$\delta_\tau = N_\tau/N_{t \gamma} \approx 10^{78}/10^{88} = 10^{-10}$ for taus are
expected if lepton matter has the same order with the critical density $\rho_e \equiv
\Omega_e \rho_c \approx \rho_c$ \cite{Roh11}. The neutrino asymmetry is $\delta_\nu =
N_\nu/N_{t \gamma} \approx 10^{88}/10^{88} = 1$ if the neutrino mass $m_\nu \approx 1$
eV. Since the baryon number and the lepton number are good quantum numbers at low
energies, excess lepton matter, consisted of electrons and neutrinos, over baryon
matter is suggested as dark matter: equal particle numbers in quarks and leptons are
crucial to the renormalization in the GWS model \cite{Glas}, which perturbatively
requires no anomaly. Based on this scheme, the quantization units of energy,
temperature, frequency, time, and distance in the universe are respectively $10^{-42}$
GeV, $10^{-30}$ K, $10^{-19}$ Hz, $10^{-43}$ s, and $10^{-33}$ cm if gravitational and
present interactions are concerned as interactions at the highest and lowest limits.
Furthermore, the proposed relation between time and gauge boson mass \cite{Roh11} is
given by
\begin{equation}
t = 1/H_e = (3/8 \pi G_N M_G^4)^{1/2} ,
\end{equation}
which provides scales in quantum cosmology as shown in Table \ref{scqc}.

At the Planck energy, a phase transition takes place from a certain group $G$ to a group $H$ and at the grand unification energy,
the group $H$ breaks down to an $SU(3)_I \times SU(3)_C$ group,
in which the $SU(3)_I$ denotes the isospin (isotope) $SU(3)$ symmetry \cite{Roh2} and the $SU(3)_C$ denotes the color $SU(3)$ symmetry \cite{Frit}:
the group chain is $G \supset H \supset SU(3)_I \times SU(3)_C$.
The Lagrangian density for the $SU(3)_C$ or $SU(3)_I$ gauge theory has the exactly same form with the Lagrangian density for gravitation.
As the energy scale decreases the $SU(3)_I$ group for isospin interactions, QWD, breaks
down to the $SU(2)_L \times U(1)_Y \rightarrow U(1)_e$ group for electroweak interactions,
known as the GWS model \cite{Glas}, and the $SU(3)_C$ group for strong interactions, QCD \cite{Frit}, analogously
breaks down to the $SU(2)_N \times U(1)_Z \rightarrow U(1)_f$ group for nuclear interactions,
known as QND \cite{Roh31}.
The $U(1)_e$ gauge theory is photon dynamics known as quantum electrodynamics (QED) and
the $U(1)_f$ gauge theory is also photon dynamics.
The DSSB of local gauge symmetry and global chiral symmetry triggers the current anomaly:
for example, the (V+A) current anomaly for weak interactions and the axial current anomaly for strong interactions.
Coupling constants for weak and strong interactions are unified around $10^{2}$ GeV or slightly higher energy rather than the order of $10^{15}$ GeV of
grand unified theory (GUT) \cite{Geor}:
$\alpha_i = \alpha_s \simeq 0.12$ according to the renormalization group analysis \cite{Roh2}.
The unification at the order of a TeV energy is consistent with the recent GUT \cite{Poma}.
Important motivations of QWD are coupling constant unification and fermion mass generation
and important concepts of QND are massive gluon and colorspin.
In terms of the Weinberg weak mixing angle $\sin^2 \theta_W = 1/4$ and the strong mixing angle $\sin^2 \theta_R = 1/4$,
coupling constant hierarchies for weak and strong interactions are respectively obtained.
Electroweak coupling constants are $\alpha_z = i_f^z \alpha_i = \alpha_i/3 \simeq 0.04$,
$\alpha_w = i_f^w \alpha_i = \alpha_i/4 \simeq 0.03$, $\alpha_y = i_f^y \alpha_i = \alpha_i/12 \simeq 0.01$, and
$\alpha_e = i_f^e \alpha_i = \alpha_i/16 \simeq 1/133$ as symmetric isospin interactions at the weak scale
and $-2 \alpha_i/3$, $- \alpha_i/2$, $- \alpha_i/6$, and $- \alpha_i/8$ as asymmetric isospin interactions:
$i_f^w = \sin^2 \theta_W$ and $i_f^e = \sin^4 \theta_W$.
Strong coupling constants for baryons are
$\alpha_b = c_f^b \alpha_s = \alpha_s/3 $, $\alpha_n = c_f^n \alpha_s = \alpha_s/4$,
$\alpha_z = c_f^z \alpha_s = \alpha_s/12$, and $\alpha_f = c_f^f \alpha_s = \alpha_s/16$
as symmetric color interactions and $- 2 \alpha_s/3 $, $- \alpha_s/2 $, $- \alpha_s/6$,
and $- \alpha_s/8$ as asymmetric color interactions:
$c_f^n = \sin^2 \theta_R$ and $c_f^f = \sin^4 \theta_R$.
The isospin factors $i^s_f = (i_f^z, i_f^w, i_f^y, i_f^e)$ or color factors introduced are
\begin{equation}
i^s_f = c^s_f = (c_f^b, c_f^n, c_f^z, c_f^f) = (1/3, 1/4, 1/12, 1/16)
\end{equation}
for symmetric interactions and
\begin{equation}
i^a_f = c^a_f = (-2/3, -1/2, -1/6, -1/8)
\end{equation}
for asymmetric interactions.  The symmetric charge factors reflect intrinsic even
parity with repulsive force while the asymmetric charge factors reflect intrinsic odd
parity with attractive force; this suggests electric-magnetic duality before DSSB. The
coupling constant chain is $\alpha_g \rightarrow \alpha_h \rightarrow \alpha_i
\rightarrow \alpha_s$ for gravitation, grand unification, weak, and strong
interactions respectively. Massive gravitons at the Planck scale produce massive
intermediate vector bosons and massive gluons at weak and strong interaction energies,
respectively. The effective coupling constant chain is $G_N \supset G_H \supset G_F
\times G_R$ for gravitation, grand unification, weak, and strong interactions
respectively: effective strong coupling constants are $G_F/\sqrt{2}= i_f g_i^2/8 M_G^2
\approx 10^{-5} \ \textup{GeV}^{-2}$ and $G_R/\sqrt{2}= c_f g_s^2/8 M_G^2 \approx 10 \
\textup{GeV}^{-2}$. Just as the gauge boson mass below the Planck energy becomes
$M_G^2 = g_f g_g^2 [A_{0}^2 - \langle \phi \rangle^2]$, the gauge boson mass below the grand
unification energy becomes
\begin{equation}
M_G^2 = M_H^2 - c_f g_s^2 \langle \phi \rangle^2 = c_f g_s^2 [A_{0}^2 - \langle \phi \rangle^2]
\end{equation}
with the grand unification mass $M_H \approx 10^2$ GeV: $M_G \simeq 300$
MeV at the QCD cutoff scale. In this scheme, all the elementary fermions such as
quarks and leptons possess three types of charges, (color, isospin, spin). Leptons are
color singlet fermions, which are governed by the isospin dynamics of QWD as well as
spin dynamics but quarks are color triplet fermions, which are governed by the color
dynamics of QCD as well as isospin dynamics and spin dynamics. Combined charges are
asymmetric configurations for both quarks and leptons to satisfy the Pauli exclusion
principle for fermions. The essential point for DSSB is that the gauge boson decreases
its mass as the gauge boson condensation increases or the charge factor (gravitational
factor $g_f$, isospin factor $i_f$, or color factor $c_f$) decreases as the energy
scale of the system decreases. The factors described above are pure color (or isospin)
factors due to color (or isospin) charges but the effective color factors used in
nuclear dynamics must be multiplied by the isospin factor $i_f^w = 1/4$ since the
proton and neutron are an isospin doublet as well as a color doublet:
\begin{equation}
c_f^{eff} = i_f^w c_f = i_f^w (c_f^b, c_f^n, c_f^z, c_f^f) = (1/12, 1/16, 1/48, 1/64)
\end{equation}
for symmetric configurations. For example, the electromagnetic color factor for the
$U(1)_f$ gauge theory becomes $\alpha_f^{eff} = \alpha/64 \approx 1/137$ when
$\alpha_s = 0.48$ at the strong scale. Note that the color mixing is the origin of the
Cabbibo angle for quark flavor mixing in weak interactions. Massless gauge bosons with
different energies are accompanied with massive gauge bosons during DSSB at different
energy scales. Table \ref{fefu} summarizes the typical features of fundamental forces.
In gravitational interactions, it is predicted that the typical cross section at a
temperature $T = 1$ GeV is $\sigma \simeq G_N^2 T^2 \approx 10^{-70} \ \textup{m}^2$
and the typical lifetime for a particle with the mass $1$ GeV is $\tau = 1/\Gamma
\simeq 1/G_N^2 m^5 \approx 10^{50}$ years, which is much longer than the proton decay
time predicted by GUT \cite{Geor}.

All the known fundamental forces in nature, strong, electroweak, and gravitational
forces on length scales from $10^{-33}$ cm to $10^{28}$ cm, may be described by local
gauge theories toward a ultimate theory for fundamental forces. Comparison between QG
and general relativity is summarized by Table \ref{cmqg}. QG is quantum theory holding
principles of gauge invariance and special relativity while Einstein's general theory
of relativity is classical theory holding principles of equivalence and general
relativity.  Newly introduced concepts in this paper are the flat universe rather than
the curved universe, DSSB rather than spontaneous symmetry breaking, gauge group
hierarchy $G \supset H \supset SU(3)_I \times SU(3)_C$, coupling constant hierarchy
$(\alpha, \alpha/3, \alpha/4, \alpha/12, \alpha/16)$ for both weak and strong
interactions, effective coupling constant hierarchy $G_N \supset G_F \supset G_R$,
effective cosmological constant $\Lambda_e = 8 \pi G_N M_G^4$, massive gauge boson
mass square $M_G^2 = M_{Pl}^2 - g_f g_g^2 \langle \phi \rangle^2 = g_f g_g^2 [A^2_0 -
\langle \phi \rangle^2]$, massless gauge bosons as Nambu-Goldstone bosons including
CMBR, the dynamical spontaneous breaking of discrete symmetries (C, P, T, and CP), QWD
as an $SU(3)_I$ gauge theory beyond the $SU(2)_L \times U(1)_Y$ weak dynamics, analogy
between QWD and QCD, Newtonian mechanics as quadrupole (tensor) interactions, the
possible modification of Einstein's general relativity, the other quantum tests such
as SIMPs in addition to WIMPs as the candidate of dark matter, the universe inflation
with the order of magnitude $10^{30}$, the baryon asymmetry $\delta_B \approx
10^{-10}$, the relation between time and gauge boson mass, the internal and external
quantization of spacetime, proton decay time much longer than one predicted by grand
unified theory, mass generation mechanism with the surface effect, etc. The gauge
boson mass $M_G \simeq 10^{-12}$ GeV associated with the cosmological constant
$\Lambda_0 \simeq 10^{-84} \ \textup{GeV}^2$ and the Hubble constant $H_0 \simeq
10^{-42}$ GeV especially indicates gauge theories for new fundamental forces
responsible for the universe expansion and CMBR.
These predictions are compatible with the recent experiments, BUMERANG-98 and MAXIMA-1.
The potential QG as a gauge theory
may resolve serious problems of GUTs and the standard model: different gauge groups,
Higgs particles, the inclusion of gravity, the proton lifetime, the baryon asymmetry,
the family symmetry of elementary particles, inflation, fermion mass generation, etc.
This scheme may also provide possible resolutions to the problems of Einstein's
general relativity or the standard hot big bang theory: the spacetime singularity,
cosmological constant, quantization, baryon asymmetry, structure formation, dark
matter, flatness of the universe, and renormalizability, etc. This work may thus
significantly contribute to the unification of fundamental forces since all the forces
in nature might be formulated in terms of local gauge theories toward the theory of
everything.

\newpage
\onecolumn

\begin{table}
\caption{\label{scqc} Scales in Quantum Cosmology}
\end{table}
\centerline{
\begin{tabular}{|c|c|c|c|c|} \hline
Scale & Time t (s) & Energy E (GeV) & Temperature T (K) & Universe Size R (cm) \\ \hline \hline
Planck $l_{Pl}$ & $10^{-43}$ & $10^{19}$ & $10^{32}$ & $10^{-3}$ \\ \hline
Weak $l_{EW}$ & $10^{-10}$ & $10^{2}$ & $10^{15}$ & $10^{14}$ \\ \hline
Strong $l_{QCD}$ & $10^{-5}$ & $10^{-1}$ & $10^{12}$ & $10^{17}$ \\ \hline
Today $l_0$ & $10^{17}$ & $10^{-12}$ & $3$ & $10^{28}$ \\ \hline
\end{tabular}
}

\vspace{1cm}

\begin{table}[h]
\caption{\label{fefu} Features of Fundamental Interactions}
\end{table}
\centerline{
\begin{tabular}{|c|c|c|c|c|} \hline
Feature & Gravity & Electromagnetic & Weak & Strong \\ \hline
\hline Gauge Boson & Graviton & Photon & Intermediate Boson &
Gluon \\ \hline Source & Spin(?) & Electric Charge & Isospin &
Color \\ \hline Coupling Constant & $\alpha_g (M_{Pl}) \simeq
0.02$ (?) & $\alpha_e \simeq 1/137$ & $\alpha_i (M_Z) \simeq 0.12$
& $\alpha_s (\Lambda_{QCD}) \simeq 0.48$ \\ \hline Gauge Boson
Mass & $10^{19}$ GeV & 0 & $10^2$ GeV & $10^{-1}$ GeV \\ \hline
Effective Coupling & $G_N \simeq 10^{-38} \ \textup{GeV}^{-2}$ & &
$G_F \simeq 10^{-5} \ \textup{GeV}^{-2}$ & $G_R \simeq 10^{1} \
\textup{GeV}^{-2}$ \\ \hline Cross Section & $10^{-70} \
\textup{m}^2$ & $10^{-33} \ \textup{m}^2$ & $10^{-44} \
\textup{m}^2$ & $10^{-30} \ \textup{m}^2$ \\ \hline Lifetime &
$10^{57}$ s & $10^{-20}$ s & $10^{-8}$ s & $10^{-23}$ s \\ \hline
\end{tabular}
}

\vspace{1cm}

\begin{table}
\caption{\label{cmqg} Comparison between Quantum Gravity and
General Relativity}
\end{table}
\centerline{
\begin{tabular}{|c|c|c|} \hline
Classification & QG & General Relativity \\ \hline \hline Exchange
Particles & massive gravitons & Massless gravitons \\ \hline DSSB
& yes & no \\ \hline Discrete symmetries (P, C, T, CP) & breaking
& no \\ \hline Monopole Confinement & yes & unknown
\\ \hline Cosmological constant & $\Lambda_e = 8 \pi G_N M_G^4$ &
$\Lambda_0 = 10^{-84} \ \textup{GeV}^2$ \\ \hline Inflation &
$10^{30}$ & no \\ \hline Matter mass generation & $\rho_m =
10^{-61} \rho_G \Theta$ & unknown \\ \hline Baryogenesis &
$10^{78}$ for $0.94$ GeV proton & no \\ \hline Leptogenesis &
$10^{81}$ for $0.5$ MeV electron & no \\ \hline Proton decay time
& $10^{57}$ s & unknown \\ \hline Singularity & no & yes \\ \hline
Universe & flat ($\Omega = 1 - 10^{-61}$) & curved \\ \hline Dark
matter & yes & unknown \\ \hline Interactions & monopole or dipole
& quadrupole (tensor) \\ \hline Renormalization & yes & no \\
\hline Relativity & special & general \\ \hline Principle & gauge
invariance & equivalence \\ \hline Free parameter & coupling
constant ($\alpha_g$) & effective coupling constant ($G_N$) \\
\hline
\end{tabular}
}

\end{document}